\documentclass[aps,prl,reprint,superscriptaddress,showpacs,amsmath,amssymb,longbibliography,lengthcheck]{revtex4-1}
\usepackage{epsfig,graphicx,float,pst-all,rotating}
\usepackage[utf8]{inputenc}
\usepackage{hyperref}
\usepackage{amssymb}
\usepackage{gensymb}

\newcommand{\be}{\begin{equation}} 
\newcommand{\ee}{\end{equation}}
\newcommand{\bea}{\begin{eqnarray}} 
\newcommand{\eea}{\end{eqnarray}}
\newcommand{\bc}{\begin{center}}
\newcommand{\ec}{\end{center}}

\begin{document}

\title{Two-particle transfer processes as a signature of shape phase transition in Zirconium isotopes} 

\author{J.A. Lay}
\affiliation{ Departamento de Física Atómica, Molecular y Nuclear,  Facultad de F\'{\i}sica, Universidad de Sevilla, Apdo.~1065, E-41080 Sevilla, Spain}
\affiliation{Instituto Interuniversitario Carlos I de Física Teórica y Computacional (iC1), Apdo.~1065, E-41080 Sevilla, Spain}
\author{A. Vitturi}
\author{L. Fortunato }
\affiliation{Dipartimento di Fisica e Astronomia "Galileo Galilei", Univ. Padova, via Marzolo, 8, I-35131 Padova, Italy }
\affiliation{INFN, Sezione di Padova, via Marzolo, 8, I-35131 Padova, Italy}
\author{Y. Tsunoda}
\author{T. Togashi}
\affiliation{Center for Nuclear Study, University of Tokyo, Hongo, Bunkyo-ku, Tokyo 113-0033, Japan}
\author{T. Otsuka}
\affiliation{Center for Nuclear Study, University of Tokyo, Hongo, Bunkyo-ku, Tokyo 113-0033, Japan}
\affiliation{Department of Physics, University of Tokyo, Hongo, Bunkyo-ku, Tokyo 113-0033, Japan}
\affiliation{RIKEN Nishina Center, 2-1 Hirosawa, Wako, Saitama 351-0198, Japan}
\affiliation{Instituut voor Kern- en Stralingsfysica, KU Leuven, B-3001 Leuven, Belgium}

\begin{abstract}
We explore two-particle transfer reactions as a unique probe of the occurence of shape coexistence in shape phase transitions. 
The (t,p) reactions to the ground state and to excited $0^+$ states are calculated for the isotope chain of even-even Zirconium isotopes starting from stable nuclei up to beyond current experimental limits.  
Two-particle spectroscopic factors derived from Monte Carlo Shell Model calculations are used, together with the sequential description of the two-particle transfer reaction mechanism.  The calculation shows a clear signature for a shape phase transition between $^{98}$Zr and $^{100}$Zr, which displays coexistence of a deformed ground state with an excited
spherical $0^+$ state. 
Furthermore, we show that there is a qualitative difference 
with respect to the case of a normal shape phase transition that can be discriminated with two-neutron transfer reactions.


\end{abstract}

\maketitle

The phases and transitions between them are prominent features of many-body systems.  The atomic nucleus, in some cases, clearly exhibits  these features.  This is a unique and precious situation because the nucleus is an isolated system, and the transition occurs as a consequence of certain changes of its ingredients rather than due to a change of external environment.  We note that because the nucleus is a finite quantal system, quantum phase transitions (QPT) may occur \cite{qpt1,qpt2}.

The phase transition can take place in different ways. Typically the change occurs as a function of control parameters as the excitation energy (i.e. the temperature in a thermodynamical framework) or the angular momentum.  But equally important are the transitions taking place in the shape of the ground state along a chain of isotopes (or isotones), where the discrete control parameter is the number of neutrons (or protons).  Meaningful order parameters systematically used in such shape evolution are, in the case of even-even nuclei, the energy of the first
$2^+$ state, the ratio $E_4$/$E_2$ and the strength of the electromagnetic E2 transition connecting ground state and the first excited $2^+$ state. However, both the excitation energies and the E2 transition depend also on the structure of other states.  It is of great and broad interest whether and how one can directly see the structure change between the ground states.

Two-particle transfer processes, e.g. reactions populating the $0^+$ states, can however provide a complementary but crucial clear-cut
signature  of the occurrence of the phase transition.  In particular, in the presence of a sharp 
or abrupt transition, one expects a sudden weakening of the usual dominance of ground-to-ground A $\rightarrow$ A+2 transitions and a corresponding abnormally strong population of one (or more) excited $0^+$ states.  The behaviour of the full pair response (as defined below) should indicate not only the occurrence of a shape phase transition, but also the nature of this transitio.

A correct description of the process implies, besides a proper reaction model,  a proper microscopic description of the nuclear wave functions and in this Letter we present novel microscopic calculations of the pair transfer process based on
nuclear structure inputs obtained within the Monte Carlo Shell Model for the chain of Zirconium isotopes \cite{tokyo}.  Before that, for the sake of clarity, we discuss schematic behaviors that can be obtained within simplified algebraically-based approaches \cite{fossion,zhang} and are meant as a guideline. 

\begin{figure}
\begin{center}
\includegraphics[width=1.\columnwidth]{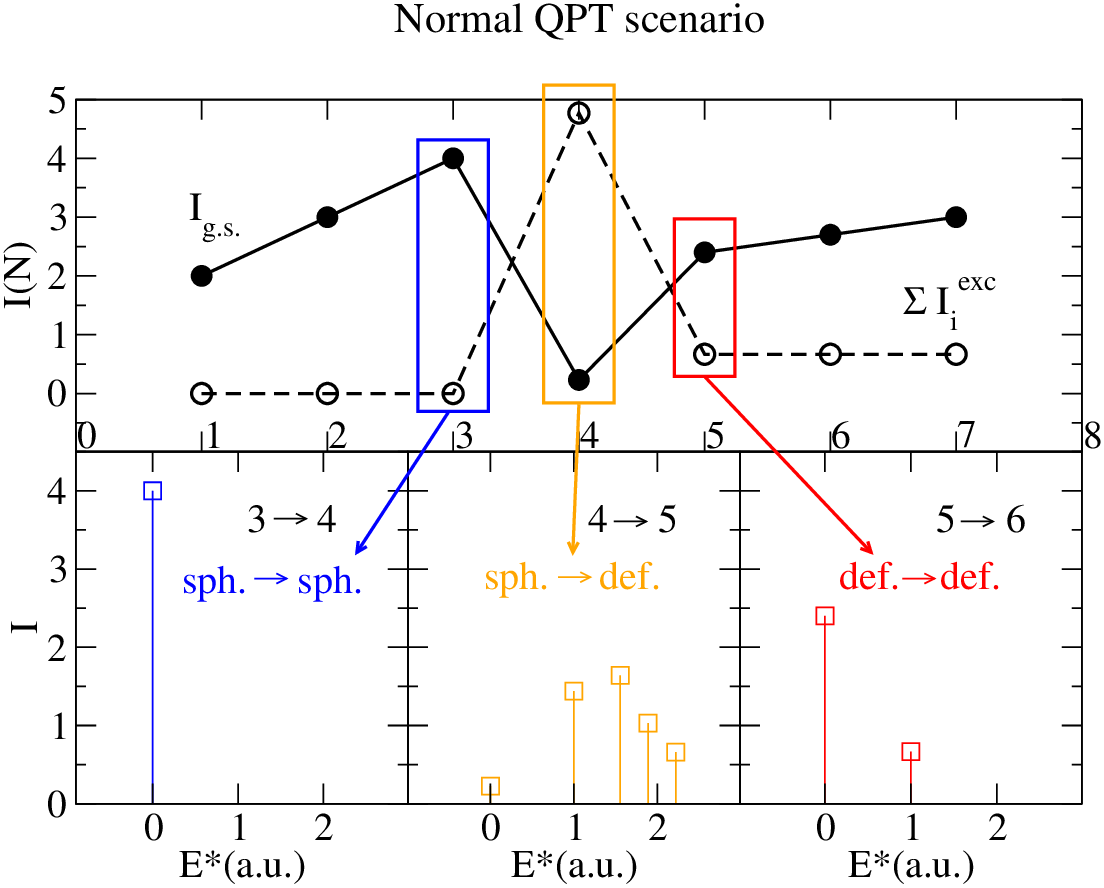}
\end{center}
\caption{ 
Upper panel: two-particle transfer intensities to the ground state (solid) and to the summed excited $0^+$ states (dashed) in the case of a sequence of isotopes making the sharp transition from U(5) to SU(3) (cf. text). In the lower panels, the full pair response spectrum is shown for the indicated cases, showing fragmentation of the strength at the critical point.
}
\label{figure1}
\end{figure} 
\begin{figure}
\begin{center}
\includegraphics[width=1.\columnwidth]
{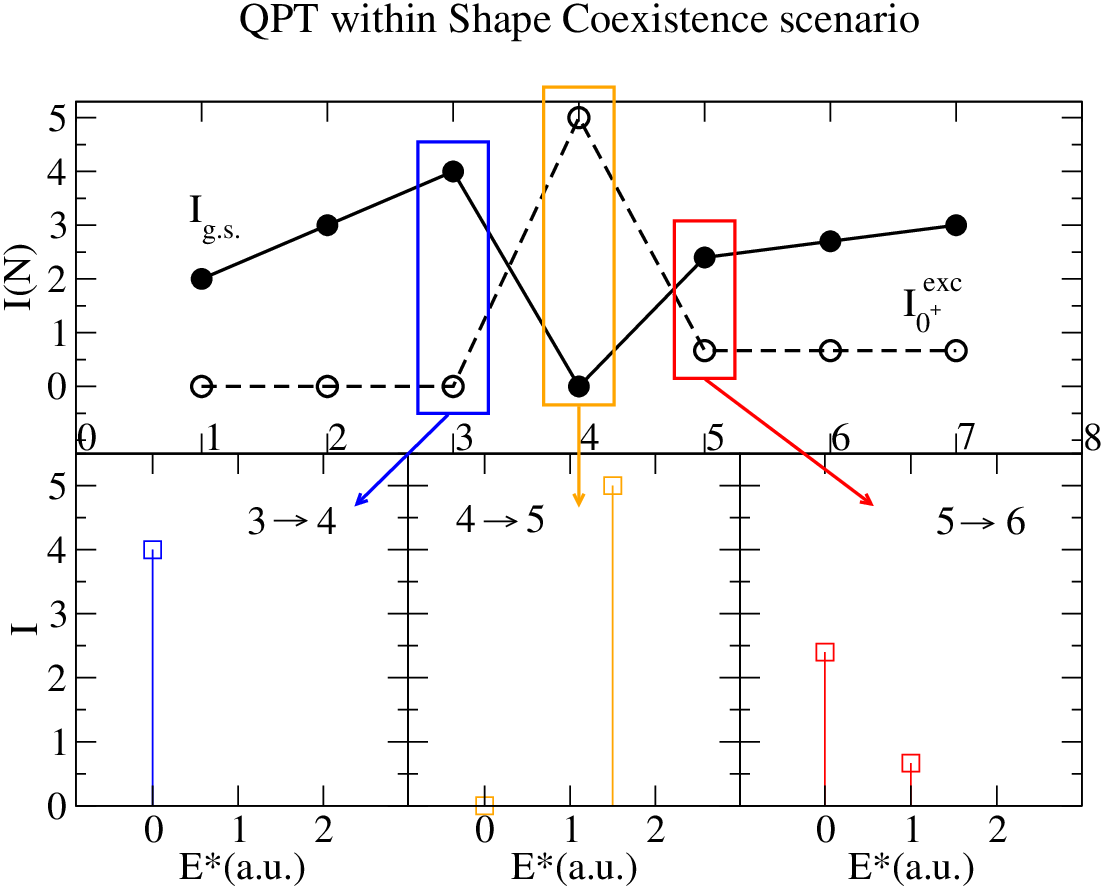}
\end{center}
\caption{ 
Two-particle transfer intensities to the ground state and to the excited $0^+$ state in the case of a sequence of isotopes in the presence of shape coexistence.  From N=4 to N=5 there is the exchange of the spherical ground-state configuration with the ``intruder" deformed configuration. In the lower panels, the full pair response spectrum is shown for the indicated cases.}
\label{figure2}
\end{figure} 

    We first consider the case of a series of isotopes described within the Interacting Boson Model (IBM), with a model hamiltonian that abruptly performs a transition from U(5) to SU(3), i.e. from a spherical behavior to the situation of deformed axial symmetry, so as to simulate the structure of the ground state of the MCSM calculation \cite{tokyo}.  As an example, considering a core with $A=90$, we can assume the U(5) hamiltonian for systems with a number of bosons, N, ranging from 1 ($^{92}$Zr) to 4 ($^{98}$Zr) and the SU(3) hamiltonian starting from N=5 ($^{100}$Zr).  The system will therefore be ``spherical" up to N=4 and ``axially-symmetric deformed" from N=5 on.
    This choice of core and valence bosons is not unique (See for example Ref. \cite{Isac}), but a different core would not qualitatively change the physical picture within the Interacting Boson Model IBM-1 model. 
    
     Within the IBM the pair creation operator, in leading order, is given just by the s$^\dagger$ operator and one can evaluate  the corresponding pair addition intensities obtained by taking the square of the matrix element connecting the ground state in system N with the ground and excited state in N+1.  These intensities are shown in the upper frame of Fig.~\ref{figure1} and display a clear ``anomaly" for the transition across the change of phase, i.e. the one connecting a ``spherical" system with a ``deformed" one.  The full pair response is shown in the  three lower frames panels of Fig.~\ref{figure1}  for a pair addition within the spherical phase (N=3$\rightarrow$4), across the phase transition (N=4$\rightarrow$5) and within the deformed phase (N=5$\rightarrow$6).  The pair strength, normally concentrated in the ground to ground transition, appears completely fragmented in correspondence of the critical point as seen in Fig.~\ref{figure1}.

A different physical scenario is that of shape coexistence, where different shape phases occur within the same nucleus at similar excitation energies~\cite{Fra04}.  We may face the situation of a (slow or rapid) progressive mixing of the spherical and deformed phases, eventually leading to the interchange of the dominant phase 
in the ground state.  Again we can have a first guess of the consequences of this situation on the pair-transfer processes within a simplified IBM-like framework. Following the idea of ref. \cite{coexistence,
enrique} we can assume for each system characterized by
N valence bosons a possible mixing of a ``spherical" state obtained within an IBM U(5) hamiltonian with another  ``deformed" $0^+$ state obtained within a SU(3) hamiltonian with N+2 bosons, microscopically originated by a 2p-2h core excitation.   In this case the pair creation operator will be given, in leading order, by $(s^{\dagger}+s)$, since we can either
add a valence-like boson or destroy the ``hole-like" boson.  Assuming a sharp transition with increasing number of particles from a fully spherical ground state to a fully deformed ground state  we obtain, still with a transition taking place passing from N=4 to N=5 (cf. inset in Fig.~\ref{figure2}), the pair transfer intensities shown in Fig.~\ref{figure2}.  As in the previous case a clear discontinuity appears at the transition point.  However, at variance with the previous case, the pair strength is always practically concentrated in a single state, without the fragmentation illustrated in Fig.~\ref{figure1} (lower central panel). Therefore, while the discontinuity just signals the occurrence of a phase transition, it is precisely the presence or the absence  of this fragmentation that characterizes the physical scenario: normal phase transition vs. phase transition driven by coexistence.

\begin{figure}[!t]
	\begin{center}
		\includegraphics[width=1.\columnwidth]{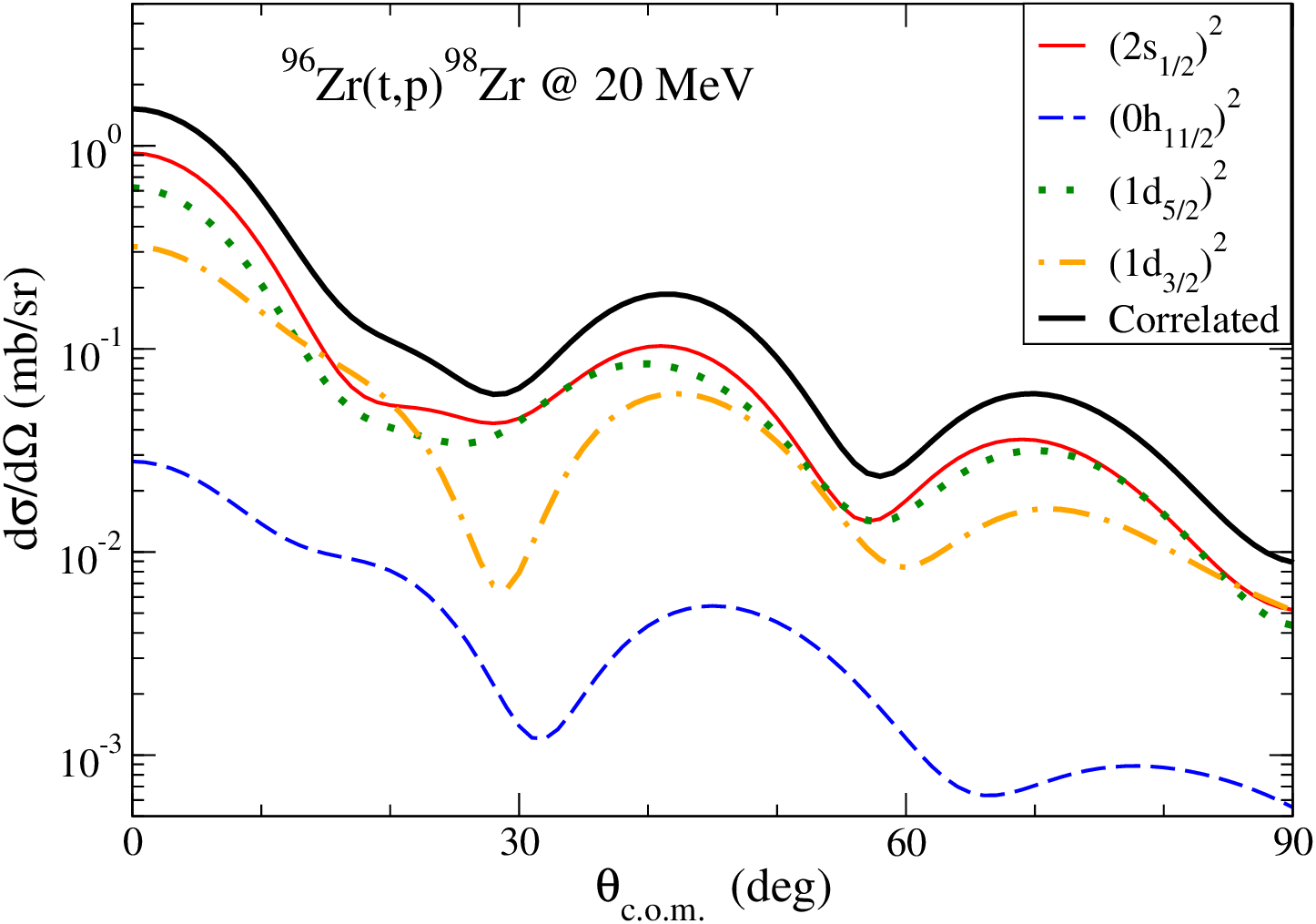}
	\end{center}
	\caption{ 
		Angular distribution for the reaction $^{96}$Zr(t,p)$^{98}$Zr at 20 MeV when the two particles are transferred in a pure configuration.  The thick solid line gives the result in the case of correlated wave function according to the value of two-particle amplitudes, the largest being given in the table I.
	}
	\label{figure3}
\end{figure}

\begin{center}
\begin{table*}[ht]
\caption{Two-particle transfer amplitudes for the different reactions connecting even-even Zr isotopes, for the most relevant single-particle orbits.  For each case the largest component is evidenced. Notice that, depending on the phase convention, an extra $(-1)^\ell$ factor should be added to these amplitudes.}
\begin{tabular}{c c c c c c c c }
\hline\hline
&90-92gs&92-94gs&94-96gs&96-98gs&98-100gs&98-100(0$^+_4$)&
100-102gs \\
\hline
1d$_{5/2}$&{\bf 0.74}&{\bf 0.86}&{\bf 0.86}&0.13&$\sim$ 0.0&0.16&0.08\\
2s$_{1/2}$&0.10&0.08&0.10&{\bf 0.90}&$\sim$ 0.0&0.16&0.05\\
1d$_{3/2}$&0.13&0.18&0.16&0.07&$\sim$ 0.0&{\bf 0.90}&0.04\\
0h$_{11/2}$&0.22&0.20&0.19&0.08&$\sim$ 0.0&0.14&{\bf 0.55}\\
\hline
\end{tabular}
\label{table1}
\end{table*}
\end{center}


\begin{figure}
\begin{center}
\includegraphics[width=1.0\columnwidth]{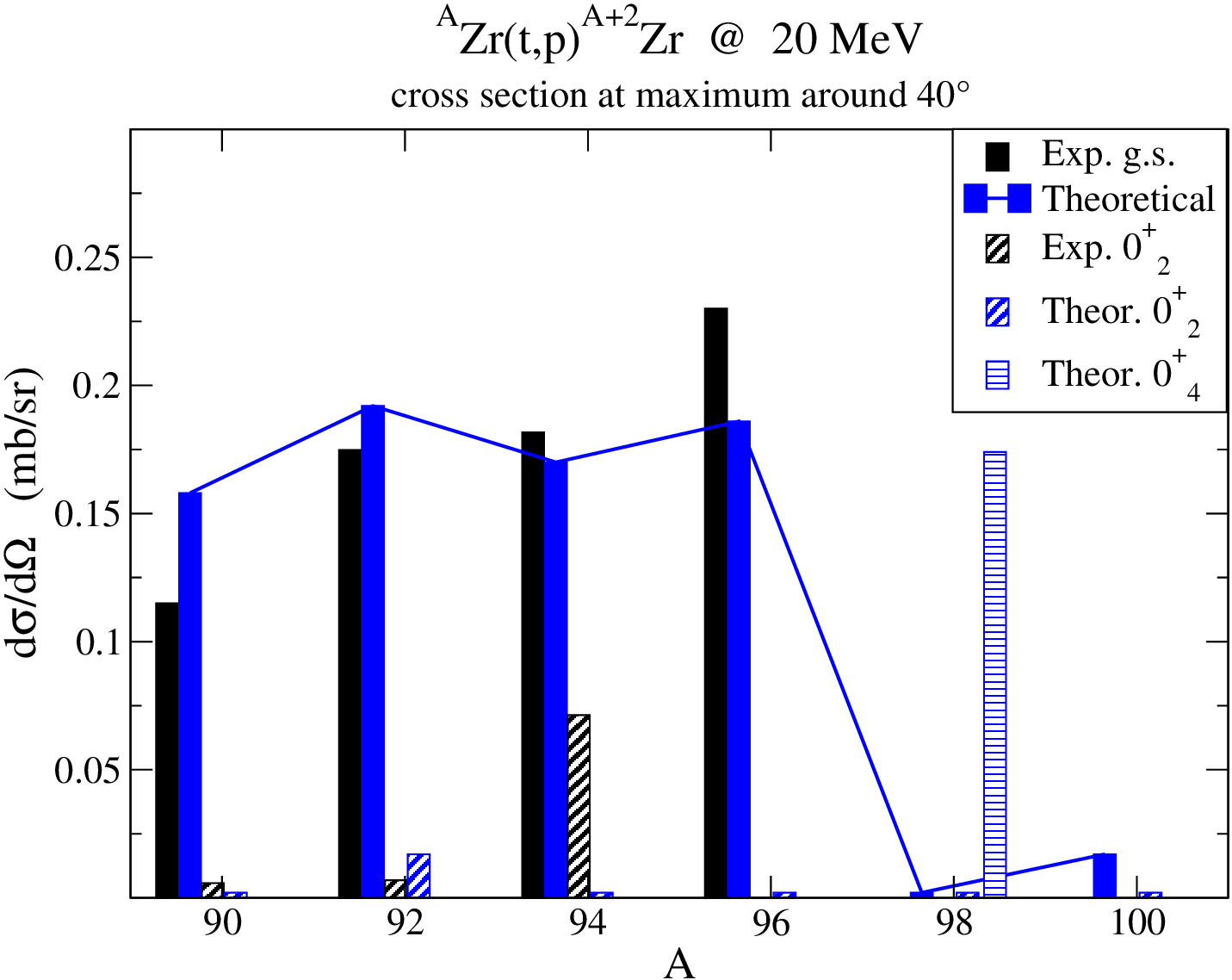}
\end{center}
\caption{ 
Differential cross sections at the maximum around 40$^\circ$ for (t,p) reactions on the different isotopes and specific final states.  Experimental values, when available, are also given~\cite{flynn}. Total cross section would be preferable although not available. Explicit values for error bars are not given in~\cite{flynn}.
}
\label{figure4}
\end{figure} 

We move now from schematic models to a fully microscopic calculation, for both reaction mechanism and structure.  We take the case of (t,p) reactions on even-A Zirconium isotopes, where experimental data at E=20 MeV are available \cite{flynn} at least for the lighter systems, i.e. up to $^{96}$Zr(t,p).  Novel interest on Zirconium isotopes has arised from the recent shell model calculations \cite{tokyo} that indicate a possible case of shape coexistence in these nuclei with a sharp transition occurring between $^{98}$Zr and $^{100}$Zr.  The situation seems to
resemble the schematic case of shape coexistence displayed in Fig.~\ref{figure2}, although the amount of the particle-hole excitations exceeds the 2p-2h picture as emphasized in \cite{tokyo}.  On the other hand, the amount of the particle-hole excitations may not matter as the normal (intruder) states are connected by the present transfer reactions to the normal (intruder) states. The deformation of Zr isotopes were already studied within the Shell Model in Ref.~\cite{Fed79} and within the IBM in Ref.~\cite{Gav19,Gar18}. Recent experimental studies can be found searching for evidences of shape coexistence in $^{98}$Zr~\cite{Sin18} and $^{96}$Zr~\cite{Kre16}.


We have therefore calculated the two-particle transfer probabilities across the phase transition up to $^{100}$Zr(t,p) to the ground and excited $0^+$ states.  In parallel with the detailed microscopic structure description described above, also the reaction process has been described in microscopic terms. In particular, we have performed second order DWBA calculations with the code FRESCO~\cite{Tho88,Tho13}. Therefore the reaction mechanism includes the ``correlated'' sequential single-particle transfer through all intermediate states in the A+1 odd system, the simultaneous transfer of the two neutrons and non-orthogonality terms. Preliminary calculations with only the sequential transfer can be found in~\cite{Vit18}. Optical models parameters have been taken as in ref.\cite{flynn} and single-particle wave functions
for the construction of the single-particle form factors have been generated within a Saxon-Woods potential adjusted to yield the proper single-particle energy.  This reaction mechanism generates a dynamical dependence on each specific orbit on which the pair is transferred. The origin of this behavior in the case of $(t,p)$ or $(p,t)$ reactions comes from the different content of $(0s)$ relative $n-n$ motions, that is associated to each two-particle configuration (cfr. how the Talmi-Moshinky brackets enter into the calculation of reaction probabilities \cite{brogliaetal}).

The transfer probabilities become therefore sensitive not only to the value of the ``global" pair strength, but also to the details of microscopic wave functions~\cite{PRClay}. This is better evidenced in Fig.~\ref{figure3} where   the cross sections associated with single-particle orbits are reported in the case of the $^{96}$Zr(t,p)$^{98}$Zr reaction.  
The collective effects in the pair trasfer process comes from the correlations present in both initial and final states that induce a coherent and constructive interference of all the sequential paths (cf. refs. \cite{brogliaetal,oertzen} and references therein).   In our case this coherence is obtained by using the 
two-particle 
spectroscopic amplitudes provided by the Monte Carlo Shell Model calculation \cite{tokyo}. 
The most important two-particle addition amplitudes are reported in the table, but also the smaller contributions from the other orbits included in the model space have been used in the reaction calculation.  The constructive effect of the residual pairing-like interaction is evidenced by the enhancement of the correlated cross section (also shown in Fig.~\ref{figure3}, full black line) with respect the single particle estimates.

  The comparison with the experimental data  is not straightforward since explicit values are not given in the only available reference~\cite{flynn}. In Fig.~\ref{figure4} that summarizes the results for the full sequence of transfer reactions, the black bars correspond to experimental data that we obtained by combining relative strengths from Figure 10 of Ref.~\cite{flynn} with the differential cross section at the first maximum (excluding $\theta=0^\circ$) given in Table 4 of the same reference.

The overall behavior of our calculations (blue bars) reproduces the experimental trend, when this is available. As expected from the amplitudes given in the table, in the case of $^{98}$Zr(t,p)$^{100}$Zr the calculation predicts a large population of the fourth $0^+$ state in $^{100}$Zr, which displays a ``spherical" behavior as the target $^{98}$Zr(gs). 
Continuing beyond the critical point, we predict again a relatively weak
population of the ground state in $^{100}$Zr(t,p)$^{102}$Zr, although the reaction connects now two deformed systems with practically the same deformation.  In this case, this is not due to small two-particle spectroscopic amplitudes (cfr. the rather large spectroscopic amplitude associated with the h$_{11/2}$ orbit in table 1), but to the reaction mechanism that does not favor the transfer of a pair to the h$_{11/2}$ single-particle level characterized by a large single-particle orbital angular momentum, 
having a small overlap with a $0s$ wavefunction in the relative $nn$ motion.

In summary, reaction calculations for $(t,p)$ processes between even-even Zr isotopes have been performed using two-particle transition amplitudes provided by state-of-the-art MCSM calculations. 
The outcome reproduces the trends of available experimental data and indicates a sharp change in the pattern of two-particle cross-sections between $^{98}$Zr and $^{100}$Zr: the ground-to-ground cross-section drops dramatically, in correspondence with a strong population of an excited $0^+$ state. This confirms that two-particle transfer reactions can be  nicely used as an additional, but crucial  probe to pinpoint the occurrence and the nature of quantum shape phase transitions. 
The higher resemblance to the IBM with configuration mixing appears to be consistent with the actual situation in the MCSM calculation. 
Finally, 
 we stress that the pair transfer reaction is a very crucial tool to look into the structure of  the wave functions of unstable nuclei, such us $^{98}$Zr and $^{100}$Zr, and it will play more central roles in the near future as the Rare-Isotope beams are becoming more intense,  being more suitable for this kind of reactions. 

\section{Acknowledgments}
The MCSM calculations were performed on the K computer at RIKEN AICS (hp150224, hp160211).
This work was supported in part by the HPCI Strategic Program (The origin of matter and the universe)
and ``Priority Issue on post-K computer" (Elucidation of the Fundamental Laws and Evolution of the Universe) from MEXT and JICFuS. This study  has been partially financed by the Consejería de Conocimiento, Investigación y Universidad, Junta de Andalucía and European Regional Development Fund (ERDF), ref. SOMM17/6105/UGR. J.A.L. acknowledges funding from the European Union’s Horizon 2020 research and innovation
programme under grant agreement N. 654002, and from the Spanish Ministerio de Economia y Competitividad and FEDER funds under Project FIS2017-88410-P.

\bibliography{trf_v3}

\end{document}